\documentclass[11pt,twoside]{atmp}

\usepackage{amsmath,amssymb}

\usepackage{graphicx}
\usepackage[all]{xy}
\usepackage{color}

\copyrightnotice{2009}{12}{1}{28} 

\begin{document}

\title[Extension of Friedmann-Robertson-Walker Theory]
{An Extension of Friedmann-Robertson-Walker Theory beyond Big Bang}

\author[Joachim Schr\"oter]{Joachim Schr\"oter}

\address{University of Paderborn, Department of Theoretical Physics\\
          D-33095 Paderborn, Germany}  
\addressemail{J@schroe.de}

\begin{abstract}
Starting from the classic Friedmann-Robertson-Walker theory with big bang it is shown that the
solutions of the field equations can be extended to negative times. Choosing a new cosmic time scale
instead of proper time one achieves complete differentiability of the scale factor and of suitable
thermodynamic quantities equivalent to pressure and energy density. Then, the singularity of big bang manifests itself only by the vanishing of the scale factor at time zero. Moreover, all solutions of the field equations are defined
for all times from $- \infty$ to $+ \infty$. In a separate chapter the horizon structure of the extended theory
is studied. Some weak assumptions guarantee that there are no horizons. Hence, the horizon problem in a strict
sence disappears. An intensive discussion of the results is given at the end of the paper.
\end{abstract}

\maketitle

\section{Introduction}

{\bf 1.1:}  Customarily, the term  Friedmann-Robertson-Walker Theory (abbreviated FRW) denotes that branch of
General Relativity which deals with homogeneous and isotropic space-times and thus is of interest for a rough description
of the universe. FRW is the starting point of the following considerations. For  the sake of definiteness let us briefly
write down the basic features of FRW as they are used later on.

\cutpage 

{\bf{FRW1:}} The space-time manifold $M^\star$ of FRW is defined by

\begin{equation} \label{1}
M^\star = N_k \times I^\star , \;  k = 0, \pm 1, 
\end{equation}

where $N_0 = \mathbb{R}^3, N_{1}$ is the 3-sphere $S^3$ and $N_{-1}$ is the 3-hyperboloid $H^3$. Moreover, $I^\star$ is either
a finite or an infinite interval of the reals. (Other possible realizations of $N_k$ are not taken into account here!).
Matter is at rest in $N_k$.

{\bf{FRW2:}} The metric $g^\star$ of FRW is given by

\begin{equation} \label{2}
g^\star = K^2 h_k - dt \otimes dt, \quad k = 0, \pm1. 
\end{equation}

The scale factor $K$ depends only on $t \in I^\star$, and $h_k$ is the Riemannian metric on $N_k$ which is induced  by the
Euclidean metric of the imbedding space $\mathbb{R}^4$ of $N_k$. The time $t$ is the proper time of the matter
fixed in $N_k$.

{\bf{FRW3:}} The field equations of FRW are the Einstein equations, the balance of energy and momentum,
and equations of continuity. These equations determine the fields $K, p^\star$ and $e^\star$, where $K$ is the scale factor,
$p^\star$ the total pressure and $e^\star$ the total energy density of the considered system. They depend only on time
$t$. The functions $p^\star$ and $e^\star$ are related to the thermodynamic state of the considered system by constitutive equations.

The Einstein equations, including a cosmological constant $\Lambda_0$, can be brought into the following form:

\begin{equation} \label{3}
2 \frac{\ddot{K}}{K} + \frac{\dot{K^2}}{K^2} + \frac{k}{K^2} - \Lambda_0 = - \kappa_0 p^\star, 
\end{equation}

\begin{equation} \label{4}
3 \frac{\dot{K^2}}{K^2} + 3 \frac{k}{K^2} - \Lambda_0 =  \kappa_0 e^\star, 
\end{equation}

\begin{equation} \label{5}
6 \frac{\ddot{K}}{K} = - \kappa_0 (3 p^\star + e^\star - \frac{2}{\kappa_0} \Lambda_0 ) . 
\end{equation}

The balance of energy and momentum reads:

\begin{equation} \label{6}
3 \frac{\dot{K}}{K} (p^\star + e^\star) + \dot{e}^\star = 0 . 
\end{equation}

The equations of continuity are not written down here because they depend specifically on the considered material system. (As is generally known, the equations (\ref{3}) to (\ref{6}) are not independent.)
 
{\bf{1.2:}} A short inspection of FRW 1 to 3 shows that with respect to geometry, FRW is almost fixed. The only
variable term is the scale factor $K$ in the metric $g^\star$. It is an unknown function of time $t$. The enormous applicability
of FRW is due to the thermodynamic side of the theory. It has its origin in the constitutive equations for $p^\star$ and $e^\star$.
These equations, together with the cosmological constant $\Lambda_0$ which has to be regarded as a
constitutive element of FRW, define the various models of the theory.

The literature about FRW is "almost infinite". Therefore I confine myself to the citing of the basic papers by Friedmann, Robertson and
Walker \cite{Fried1, Fried2, Rob1, Rob2, Walk1}
and to some monographs from which I derived special advantage \cite{Wein1, Rind1, Boer1, Kolb1, Haw1, Goe1}. 
Many others could be cited as well.

{\bf{1.3:}} In what follows, attention is mainly focused on that class of models of FRW which exhibit the phenomenon
called big bang. 

With respect to the scale factor $K$, this class is characterised by the following four conditions.

\begin{description} 
\item{\bf B1:} \parbox[t]{.9\textwidth}{$K(t) > 0$ for $t \in I^\star$.}
\item{\bf B2:} \parbox[t]{.9\textwidth}{There is a time $t_1 \in I^\star$ for which $\dot{K} (t_1) > 0$.}
\item{\bf B3:} \parbox[t]{.9\textwidth}{$I^\star = ] 0, T [$ where $T$ is finite or infinite.}
\item{\bf B4:} \parbox[t]{.9\textwidth}{$\lim_{t\to0} K (t) = 0$.}
\end{description}

Condition B4 together with its implications relating to $p^\star$ and $e^\star$ constitutes what usually is called big bang.
(For the motivation of B1 to B4 cf. e.g. the cited literature.)

In order to have a simple and convenient notation I use in the following  the term FRW in a somewhat restricted sense:
It is characterised by FRW1 to FRW3 together with the conditions B1 to B4. This means that in this paper we are
(almost) always concerned within the frame of FRW1 to 3 with cosmological models exhibiting big bang. Only in
Section 2.6 some remarks are added about other models which are connected to an extension of FRW.

{\bf{1.4:}} It is well known that big bang is not an event in the sense of General Relativity, i.e. there is 
no point $x \in M^\star$ which corresponds to a natural phenomenon we call big bang. Nevertheless, the question
is legitimate if it is possible to change the theory in such a way that big bang is describable as an event, i.e.
as a point of some Lorentz manifold. Clearly, as big bang is theoreticly understood to be a singularity there is
a problem to what extend it can  be regularized. By these remarks the problem we are dealing with in this paper is briefly outlined.

\setcounter{page}{4}

\noindent

\section{The extended Friedmann-Robertson-Walker Theory}

{\bf{2.1:}} In a first step  of extending FRW, a proposition is formulated and proved which is the key toward
EFRW. For this purpose we need some

{\bf {Notation:}} 1.
Let $K$ be the scale factor of FRW which is defined on $I^\star = ]0, T[$.

Then define $R_\pm$ by

\begin{equation} \label{7}
R_\pm (t) = \left\{  
\begin{array}{rl}
K(t),& t \in I^\star, \\
\pm K (-t), &t \in - I^\star.  
\end{array} \right.
\end{equation}

Moreover let the functions $p$ and $e$ be defined by

\begin{equation} \label{8}
p (t) = p^\star(|t|), \quad e(t) = e^\star (|t|),
\end{equation}

where $t \in I^\star \cup - I^\star$. (The notation implies that $t$ is the same parameter as in FRW!) \\
2. Since in what follows we are mainly concerned with $R_-$, notation is simplified by omiting the signs
$+$ or $-$. If necessary, we write $R = R_+$ or $R = R_-$.

Now the basic theorem can be  stated thus.

{\bf{Proposition:}}
Let $K, p^\star, e^\star$ be a solution of the equations (\ref{3}) to (\ref{6}). Then both \\ $R = R_+, p, e$ and
$R = R_-, p, e$ are solutions of the equations

\begin{equation} \label{9}
2 \frac{\ddot{R}}{R} + \frac{\dot{R^2}}{R^2} + \frac{k}{R^2} - \Lambda_0 = - \kappa_0 p, 
\end{equation}

\begin{equation} \label{10}
3 \frac{\dot{R^2}}{R^2} + 3 \frac{k}{R^2} - \Lambda_0 =  \kappa_0 e, 
\end{equation}

\begin{equation} \label{11}
6 \ddot{R} +  \kappa_0 R (3 p + e - \frac{2}{\kappa_0} \Lambda_0 ) = 0,
\end{equation}

\begin{equation} \label{12}
3 \frac{\dot{R}}{R} (p + e) + \dot{e} = 0 
\end{equation}

for all $t \in I^\star \cup - I^\star$.
The equations (\ref{9}) to (\ref{12}) have the same interdependences as (\ref{3}) to (\ref{6}).

{\bf Proof:} Since for $t \in I^\star$ the relations $R(t) = K(t), p(t) =  p^\star(t)$ and 
$e(t) = e^\star(t)$ hold, the equations (\ref{9}) to (\ref{12}) are identical with (\ref{3}) to (\ref{6}).
If $t\in - I^\star$, one finds that

\begin{equation} \label{13}
R (t) = \pm K (-t), \; \dot{R}(t) = \mp \dot{K} (-t), \; \ddot{R}(t) = \pm \ddot{K} (-t) 
\end{equation}

and

\begin{equation} \label{14}
p (t) = p^\star(-t), \; e(t) = e^\star (-t), \;  \dot{e}(t) = - \dot{e^\star} (-t). 
\end{equation}

Now inserting (\ref{13}) and (\ref{14}) into (\ref{9}) to (\ref{12}) one again finds by a simple calculation
that (\ref{3}) to (\ref{6}) hold. Thus the proposition is seen to be true.

Some special cases of the proposition can be found in the literature.

Since the domain of a solution $R = R_{\pm}, p$ and $e$ of (\ref{9}) to (\ref{12})  is the set $I^\star
 \cup - I^\star$, the functions
$R, p, e$ are not defined for $t = 0$. Thus the problem sketched in Section 1.4 now reads:\\
1. Is it possible to define the triple $(R, p, e)$ or some equivalent also for $t = 0$?\\
2. Are the considered functions, if definable for $t = 0$, also continuous or even differentiable at this point?\\
This problem will be treated for $R = R_-$ in the Sections 2.2 to 2.5. Some remarks concerning $R_+$ are added in 
Section 2.6.

{\bf 2.2: } In this subsection a function $R$ is considered which is the first component of a solution
$(R, p, e)$ of (\ref{9}) to (\ref{12}). More specifically, we are interested in the behaviour of $R$ in a
neighbourhood of $t=0$. From the definition of $R$ we arrive at the following two

{\bf Consequences:} 1. Since by assumption only such models of FRW are taken into account for which 
$\lim_{t\to0^+} K(t) = 0$ we find that for each $R$ the relation

\begin{equation} \label{15}
\lim_{t\to0} R (t) = 0
\end{equation}

holds for arbitrary limits. Hence, if we define $R (0) = 0$, the function $R$ has the domain $I = ]-T, T [$ and is continuous.\\

2. By definition of $R$ (with $R = R_-)$ the relation $\dot{R}(-t) = \dot{R}(t)$ holds. Therefore from equation
(\ref{10}) one concludes that

\begin{equation} \label{16}
\lim_{t\to0} \dot{R} (t) := a_0 < \infty
\end{equation}

exactly if

\begin{equation} \label{17}
\lim_{t\to0} e(t) R (t)^2 := b_0 < \infty.
\end{equation}

3. By (\ref{7}), (\ref{8}) and (\ref{15}) the phenomenon of a big bang in FRW is replaced by a big crunch  bang in EFRW.
This means, if (\ref{17}) is valid, the function $R$ is continuously differentiable in the whole domain $I$ of $R$.
However, as we shall see, condition (\ref{17}) is not fullfilled by realistic constitutive equations for $e$.

{\bf 2.3:} The question how the thermodynamic quantities $p$ and $e$ behave for $t \rightarrow 0$ can only be 
answered if some assumptions about the constitution of the universe for $t \rightarrow 0$ are imposed.
The generally accepted supposition is the following:

The universe is radiation dominated in a neighbourhood of $t = 0$ in the sense that the influence of ponderable matter
is asymptotically negligible if $t \rightarrow 0$. In other words, the model of a radiation dominated universe gives
an asymptotically correct description of the real situation in the vicinity of $t=0$

This assumption has the following

{\bf Consequences:} 1. Let $\eta$ denote the density of photons. Then $e = \eta$ and $p = \frac{1}{3} \eta$,
so that the solution of (\ref{12}) reads

\begin{equation} \label{18}
\eta = A R^{-4}, A = \eta(t_0) R(t_0)^4.
\end{equation}

For the sake of simplicity let us assume that $\Lambda_0 = 0$. Then the solution of (\ref{10}) and (\ref{18}) is given by

\begin{equation} \label{19}
R (t)^2 = (\frac{4}{3} \kappa_0 A)^{\frac{1}{2}} |t| - k t^2, \; k = 0, \pm 1
\end{equation}
 
and

\begin{equation} \label{20}
R (t) = \pm (R(t)^2)^{\frac{1}{2}},\; \mbox{if} \; \pm t \geq 0.
\end{equation}

The result is well-known for $t >0$.\\
2. If one considers the specific photonic volume $\omega = \eta^{-1}$ instead of $\eta$, then by (\ref{18}) one finds that
$\omega = A^{-1} R^4$. Hence $\omega$ is continuous at $t = 0$ but not differentiable because $R$ is not differentiable.
The equations (\ref{9}) and (\ref{10}) now take the form

\begin{equation} \label{21}
(2 R \ddot{R} + \dot{R}^2 + k - \Lambda_0 R^2) R^2 = -\frac{1}{3} \kappa_0 A 
\end{equation}

\begin{equation} \label{22}
(3 \dot{R}^2 + 3k - \Lambda_0 R^2) R^2 =  \kappa_0 A. 
\end{equation}

3. The equations (\ref{21}) and (\ref{22}) are used to determine the asymptotic behaviour of $\dot{R}$ and $\ddot{R}$. Since,
by supposition the relation (\ref{15}) holds, one concludes from (\ref{22}) that

\begin{equation} \nonumber
\lim_{t\to0} \dot{R}^2 R^2 = \frac{\kappa_0}{3} A > 0.
\end{equation}

Hence

\begin{equation} \label{23}
\lim_{t\to0} \dot{R} |R| = \left(\frac{\kappa_0}{3}\right)^{\frac{1}{2}} A^{\frac{1}{2}} 
\end{equation}

or
\begin{equation} \label{24}
\dot{R} = {\mathcal{O}} (|R|^{-1}) \quad \mbox{for} \; t \rightarrow 0.
\end{equation}

The derivative $\dot{R}$ diverges like $|R|^{-1}$. From (\ref{21}) we see that

\begin{equation} \label{25}
\lim_{t\to0} R^3 \ddot{R} =  -\frac{1}{3} \kappa_0 A^{-1} < 0 
\end{equation}

or

\begin{equation} \label{26}
|\ddot{R}| = {\mathcal{O}}   (|R|^{-3})\quad  \mbox{for}\; t \rightarrow 0.
\end{equation}

Hence the second derivative $\ddot R$ diverges like $|R|^{-3}$.

{\bf2.4:} The equations (\ref{23}) to (\ref{26}) suggest asking whether it is possible to obtain differentiability of $R$ by a
transformation of the time parameter $t$.

{\bf 2.4.1:} This is indeed possible as the following considerations show.

Let $f$ be a bijective function which is two times continuously differentiable and satisfies 
the relation $t = f(\tau) = -f(-\tau)$. Now define $S$ by $S (\tau) = R (f(\tau))$. Then it follows 
from (\ref{13}) that 

\begin{equation}  \label{100}
 S(\tau) = -S(-\tau),
\end{equation}  

\begin{equation} \label{101} 
 \dot{S}(\tau) = \dot{S}(-\tau),\; \ddot{S}(\tau) = -\ddot{S}(-\tau) \quad \mbox{for} \;  \tau \neq 0.
\end{equation}

Therefore the equation $ S(0) = 0 $ is a necessary condition. Moreover, if $ S $ is two times continuously differentiable 
in $I$, it is also necessary that $ \ddot{S}(0) = 0  $ holds. 
Now let us consider sufficient conditions for $ S $ to be differentiable. By definition of $ S $ we find for $ \tau \neq 0$ that

\begin{equation} \label{27}
\dot{S}(\tau) = \dot{R} (f (\tau)) \dot{f} (\tau),
\end{equation}

\begin{equation} \label{28}
\ddot{S}(\tau) = \ddot{R} (f (\tau)) (\dot{f}(\tau))^2 +   \dot{R} (f(\tau)) \ddot{f} (\tau).
\end{equation}

From (\ref{24}) and (\ref{27}) one concludes that $\dot{S} (0)$ exists if

\begin{equation} \label{29}
\dot{f}(\tau) = {\mathcal{O}} (|S (\tau)|) \quad \mbox{for} \; \tau \rightarrow 0.
\end{equation}

A similar result holds for the second derivative. From (\ref{26}) to (\ref{28}) one derives that $\ddot{S} (0)$ exists  
and that $\ddot{S}(0) = 0$ if

\begin{equation} \label{30}
\dot{f}(\tau) = o(|S (\tau)|^{\frac{3}{2}}), \; \ddot{f} (0) = o(|S (\tau)|) \quad \mbox{for} \; \tau \rightarrow 0.
\end{equation}

{\bf 2.4.2:} These general results now are illustrated by a significant example. Let us consider the radiation dominated
era with $\Lambda_0 = 0$. Then according to (\ref{19}) and (\ref{20}) we obtain with $B: = \frac{4}{3} \kappa_0 A$ that

\begin{equation} \label{31}
R (t) = \pm (B |t| - k t^2)^{\frac{1}{2}}, \quad \pm t \geq 0.
\end{equation}

Now define $t = f(\tau): = \alpha \tau^n$ where $n$ is an odd positiv integer and $\alpha > 0$. Inserting $f(\tau)$ into (\ref{27}) and (\ref{28}) one obtains

\begin{equation}\label{32} 
 \dot{S}(\tau) = |\tau|^{\frac{n}{2}-1} F_1 (\tau),\;  
\ddot{S}(\tau) = \pm |\tau|^{\frac{n}{2}-2} F_2 (\tau), 
\end{equation}

where $F_1$ and $F_2$ are continuously differentiable functions and where

\begin{equation} \label{33}
F_1 (0)=  \frac{n}{2} \alpha^{\frac{1}{2}} B^{\frac{1}{2}}, \; F_2(0) = \frac{n}{4} (2n-3) \alpha^{\frac{1}{2}} B^{\frac{1}{2}}.
\end{equation}

From (\ref{32}) we draw the conclusion that the transformed scale factor $S$ is two times continuously differentiable if
$n \geq 5$ and that the equations $ S(0) = 0, \dot{S}(0) = 0$ and $\ddot{S}(0) = 0$  hold.

{\bf 2.4.3:} The example $f(\tau) = \alpha \tau^n$ of a time transformation shows the special feature that the proper
time $t$ is totally replaced by time $\tau$. But proper time $t$ has proved its worth far from $t=0$ or $\tau = 0$ respectively.
Therefore one should define $f$ for large $t$ by $t = f (\tau):= \tau$. Qualitatively the function $f$ should have the
form as illustrated in Figure 1.

\begin{center}
\includegraphics[scale=0.7]{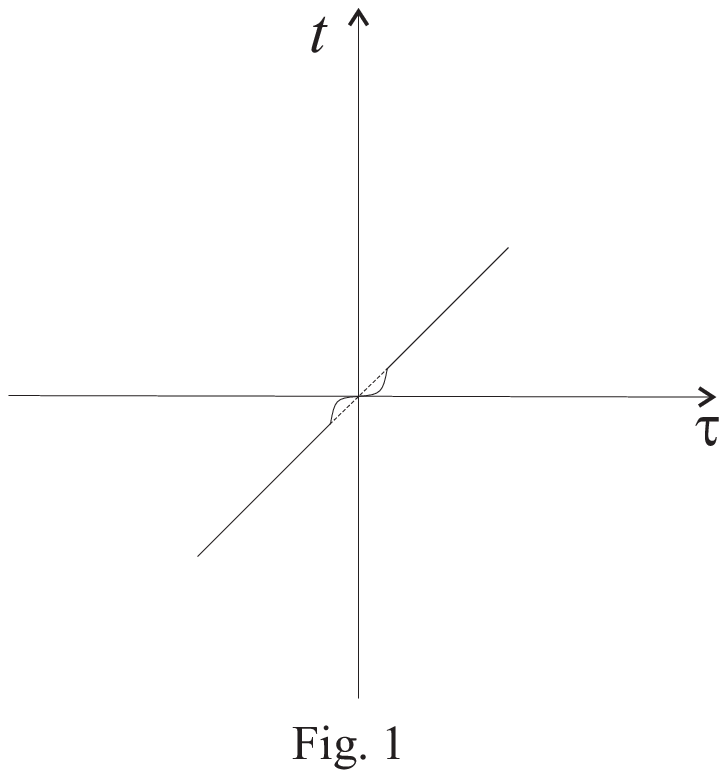 }
\end{center}

In case that $T$ is finite it is useful to change the function $f$ defined on $[- T, T]$ in a neighbourhood of $- T$ and of
$T$ in the same way as it was done in a neighbourhood of $\tau = 0$. Then qualitatively $f$ looks as in Figure 2.

\begin{center}
\includegraphics[scale=0.7]{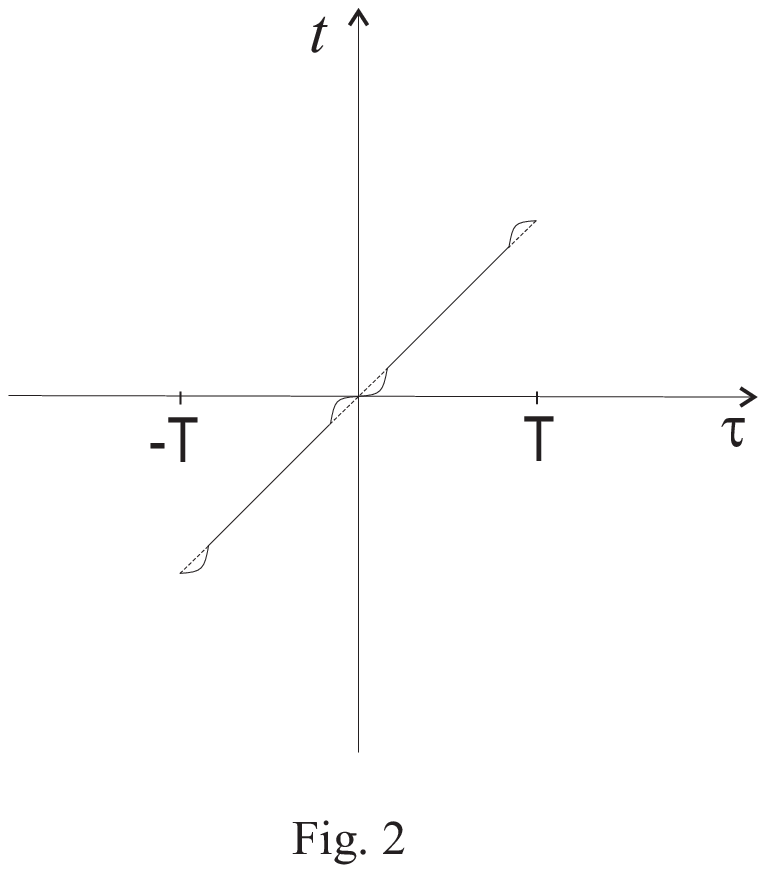}
\end{center}

In this case $f$ satisfies the conditions

\begin{equation} \label{35}
\dot{f}  (\pm T) = 0 , \quad  \ddot{f} (\pm T) = 0.
\end{equation}

If $f$ is defined this way, the solutions of the field equations defined in the intervals $[(2 z-1) T, (2z + 1) T], 
z \in \mathbb{Z}$ can be connected
so that an infinitely periodic scale factor results.

{\bf 2.4.4:} Special attention with respect to differentiability and continuity has to be paid to the constitutive
quantities $p$ and $e$. As already remarked in Section 2.3 (Consequence 2.), a transformation of the time scale 
$t = f(\tau)$ is not sufficient for removing the divergence of $p$ and $e$ at $t=0$ or $\tau =0$. The pressure $p$ and the
energy density $e$ are not suitable quantities to describe big crunch bang smoothly! Which quantities are suitable?
In order to find an answer let us consider $\bar p(\tau) = p (t)$, $\bar e(\tau) = e (t), t = f(\tau)$, and 
in addition the quantities $\psi = \bar p^{-1}$ and $\omega = \bar e^{-1}$. Moreover, since $S$ is bijective in a 
neighbourhood of
$\tau =0$ we obtain $\psi (\tau) = \hat{\psi} (S(\tau))$, $\omega (\tau) = \hat{\omega} (S(\tau))$, and arrive at the following

{\bf Result}: The constitutive quantities $\psi$ and $\omega$ are suitable for a smooth description of big crunch bang
if the derivatives of $\hat{\psi}$ and $\hat{\omega}$ up to the second order are continuous in a neighbourhood of
$S (\tau)=0$.

{\bf 2.4.5:} If the time transformation $f$ is different from $id$, i.e. if $t=\tau$ does not hold throughout,
 the field equations (\ref{9}) to (\ref{12}) have to be modified. This can be achieved most easily by inserting
$R (t) = S(\tau)$, $p(t) = \bar p (\tau)$ and $e(t) = \bar e(\tau)$ together with 
$\tau = f^{-1} (t) $ into (\ref{9}) to (\ref{12}).
The result is given in  Section 2.5 formula (\ref{38}).

{\bf 2.5:} The heuristic considerations of the Section 2.1 to 2.4 now lead to a compact description of EFRW.

{\bf 2.5.1:} The basic properties of EFRW are condensed in the following four conditions

{\bf EFRW 1:} The manifold of events $M$ is given by $M = N_k \times I, k = 0, \pm 1$, where 
$N_0 = \mathbb{R}^3, N_1$ is the 3-sphere and $N_{-1}$ is the 3-hyperboloid. Moreover, $I = ] - T, T[$. There
is a global chart on  $I$ the coordinate function of which is the identy $id$. The time parameter is denoted $\tau$
and is called the (new) cosmic time scale.

{\bf EFRW 2:} 1. The metric $g$ is defined on $M$. It has the form

\begin{equation} \label{36}
g = S^2 h_k - \dot{f}^2 d \tau \otimes d \tau .
\end{equation}

2. The tensor $h_k$ is the Riemannian metric on $N_k$ which is induced by the Euclidean metric of the 
imbedding space $\mathbb{R}^4 $ of $N_k$.\\ 
3. The scale factor $S$ depends only on $\tau$. It is of class $ C^r, r\ge 2.$ \\
4. The time parameter $\tau$ and the proper time $t$ of the matter fixed in $N_k$ are related by an equation
$t= f(\tau)$, where $f$ is an increasing function of class $C^l, l\ge 3$ which is given as a constitutive element satisfying the condition 

\begin{equation} \label{37}
f(-\tau) = -f (\tau) .
\end{equation}

Moreover the inverse function $f^{-1}$ is also of class $C^l$ for $t \neq 0$.

{\bf EFRW 3:} 1. The field equations are the equation (\ref{9}) to (\ref{12}) where the time $t$ is  transformed
by $t = f(\tau)$. The constitutive quantities $\bar p$ and $\bar e$ defined by $\bar p (\tau) = p (f (\tau)), 
\bar e (\tau) = e (f(\tau))$
can be replaced by suitable quantities $\psi,\omega$ such that the field equations  contain only the functions $S, \psi, \omega$ which are
defined for each $\tau \in I$. If $\bar p \not= 0$ and $\bar e \not= 0$ the quantities $\psi = \bar p^{-1}$ 
and $\omega = \bar e^{-1}$ are
suitable. In this case the field equations read:

\begin{equation}\label{38}
\begin{array} {lll}
\psi(2 S q \ddot{S} + q \dot{S}^2 - 2 S \dot{q} \dot{S} + k q^3 - \Lambda_0 q^3 S^2)  = -\kappa_0 q^3 S^2 \\ [0,3cm]
\omega(3 \dot{S}^2 + 3 k q^2 - \Lambda_0 q^2 S^2)  = \kappa_0 q^2 S^2 \\  [0,3cm]
6 q \omega \psi \ddot{S} = - \kappa_0 q^3 S (3 \omega +\psi) + 6 \psi \omega \dot{q} \dot{S} + 2 \Lambda_0 q^3 S \omega \psi\\ [0,3cm]
3 \omega^2 (\omega+\psi) \dot{S} - \psi \omega S \dot{\omega} = 0
\end{array}
\end{equation}

where $q=\dot{f}$.\\
2. The  functions $S, \psi, \omega$ satisfy the following relations

\begin{equation} \label{39}
\begin{array}{ll}
S(\tau) = - S(- \tau), \quad \psi (\tau) = \psi (- \tau), \quad \omega (\tau) = \omega (-\tau)\quad
 \mbox{for all} \; 
\tau \in I,\\ [0,5cm] S(\tau) > 0  \quad \mbox{for} \quad \tau >0 \quad \mbox{and} \quad \dot{S}(0) = 0.
\end{array}
\end{equation}

Because of $ \dot{R}(0) \ne 0$ and (\ref{27}) it follows from (\ref{39}) that $\dot{f}(0) = 0$.

{\bf EFRW 4:} Outside a time interval $[-\epsilon, \epsilon], \epsilon >0$, the formulation of EFRW in terms
of  $t, R, p, e$ is equivalent to that in terms of $\tau, S, \psi, \omega$. In this case $t$ is given by two charts on $I$,
one for $t >0$ the other for $t <0$ (cf. Subsection 2.5.3.1).

{\bf 2.5.2.:} The strategy for solving (\ref{38}) and (\ref{39}) is extremely simple. It can be achieved in four steps.\\
1. Let the constitutive equations for $p^\star$ and $e^\star$ be given. Then, first of all one solves the equations (\ref{3}) to
(\ref{6}) from FRW with the initial condition $lim_{t \to 0} K (t) = 0$.\\
2. If $K, p^\star$ and $e^\star$ are known one defines the functions $R, p$ and $e$ by (\ref{7}) and (\ref{8}). They satisfy the 
equations (\ref{9}) and (\ref{12}).\\
3. By a time transformation $t = f(\tau)$ and the choice of suitable quantities, e.g. $\psi = p^{-1}, 
\omega = e^{-1}$
a solution $S, \psi, \omega$ of (\ref{38}) and (\ref{39}) is obtained.\\
4. If one is only interested in times unequal zero the two descriptions in terms of $t, R, p, e$ and in terms of
$\tau, S, \psi, \omega$ are equivalent. For practical purposes the time scale $t$ is generally more comfortable than the
scale $\tau$.

{\bf 2.5.3:} In this subsection some remarks are added to illustrate the results of EFRW.

{\bf 2.5.3.1:} Since $R (0) = 0$, the metric $g$ written down in coordinates with time $t$ (i.e. $g$ of the form
(2)) is singular at $t=0$. This means $g=dt \otimes dt$ and det $g=0$. After the  time transformation $t=f(\tau)$
one obtains $dt = \dot f(\tau) d\tau$ so that $dt = 0$ for $\tau=0$. Hence $g=0$ for $\tau=0$, because
$S(0) =0$ and $\dot{f}(0) = 0$. These properties show that the time scales $t$ and $\tau$ are not compatible
at $t=0$ or $\tau =0$ respectively. 
What is the right scale?

On mathematical grounds, $\tau$ is preferable because $S (\tau) = R (f(\tau))$ is differentiable for all
$\tau \in I$, whereas $R(t)$ is not differentiable at $t=0$. Expressed in geometrical terms this means, the identity
function $id$ is a global time chart on $I$, whereas the time $t=f(\tau)$ is not given by a global time chart because 
$f^{-1}$ is not differentiable at $t=0$, but only continuous throughout. Thus, if we introduce the charts $f_\pm$ by $f_\pm (\tau) = f(\tau)$ for
$\pm \tau > 0$ time $t$ is seen to be given by two charts $C^l$-compatible ($l\ge 3$) with $\tau$. Both these charts can
sometimes be used for convenient calculating.  The change of the cosmic time scale from $t$ to $\tau$ is one of
the essential features of EFRW. Nevertheless, there is one special point: $g=0$ for $\tau = 0$. We shall
come back to it in Section 4.1.4. Physical arguments in favour of $\tau$ are given in Section 4.2.

{\bf 2.5.3.2:} As already pointed out in Section 1.2, the constitutive equations for $p^\star$ and $e^\star$ and the
possible values of $\Lambda_0$ define the different models of FRW and consequently the models of EFRW. This implies
that all results obtained in  FRW (in the sense of Section 1.3) are also valid in EFRW if the asymptotic behaviour of
$R$ (cf. (\ref{23}) to (\ref{26})) is guaranteed. This implies $\lim_{t\to 0^+} K (t)=0 $.
But it is well known that there are constitutive scenarios $p^\star, e^\star, \Lambda_0$ such that $K$ does not satisfy this
condition (cf. e.g.\cite{Rind1}).

As usual, if cosmological problems are treated within the frame of FRW the constitutive equations cannot be specified
by only one function of the thermodynamic state for the whole time interval $I^\star$, rather one uses different 
constitutive equations for different subintervals of $I^\star$. EFRW exhibits the same features. But one can even go a step
further. The constitutive scenario for negative times $\tau$ is in EFRW totally determined by the scenario for
positive $\tau$. If there are reasons to impose other constitutive relations for $\tau < 0$ this can be done very
easily. One only has to set up two models of FRW and glue them together according to the rules developed in the
previous sections. Up to now, no such reasons are known!

{\bf 2.5.3.3:} It was already mentioned at the end of Section 2.4.3 that models of FRW with finite time intervals
$I^\star$ can be glued together if the time scales $t$ and $\tau$ are related by a function $f$ as qualitatively
illustrated in Figure 2. This means that each model of FRW with finite $I^\star$ generates a periodic model of EFRW,
i.e. a model with infinite $I$. As long as there is no reason for considering only a finite part of a periodic
model all models in EFRW are defined for all $\tau \in I= \mathbb{R}$.

{\bf 2.5.3.4:} Summing up, EFRW is characterized by the  following features: the  new cosmic time scale $\tau$,
the use of suitable constitutive quantities, e.g. $\psi, \omega,$ and the infinite time interval $I = \mathbb{R}$.

{\bf 2.6:} The considerations of the Sections 2.2 to 2.5. refered only to $R:=R_-$ and $S=R(f)$. But the results
there obtained are almost all valid also for $R_+$ and $S_+ = R_+ (f)$. One of the differences between
$R:=R_-$ and $R_+$, and $S$ and $S_+$ respectively, is that $R$ and $S$ are monotone functions in an interval containing big crunch bang, hence invertible, whereas
$R_+$ and $S_+$ are not. In the next chapter this property of $R$ will be exploited.

Nevertheless, $R_+$ comes into play if in FRW scenarios without big bang are considered. A nontrivial example of this
kind was given by Lessner (cf. \cite{Less1}). I will come back to these aspects of EFRW in a subsequent paper.

\section{Horizons}

\subsection{General Remarks}
{\bf 3.1.1:}
The concept of horizon is well established in  FRW (cf. \cite{Wein1, Rind1, Rind2, Boer1, Kolb1, Haw1, Goe1}).
In this chapter it will be introduced also in EFRW. More precisely, we are looking for a common concept in FRW and EFRW without
altering the results in FRW. For this purpose we need to fix some

{\bf Notation:} 1. Let the manifold of FRW be $M^\star = N_k \times I^\star$ and let that of EFRW be $M = N_k \times I$.
Then in both these cases the time coordinate is chosen to be proper time $t$. This choice is possible because we want to study intgrals and for these purposes $t$ is more comfortable than time $\tau$.
Moreover, if $N_k$ has to be coordinatized we take the triple $(\chi, \vartheta, \varphi)$ where $\chi$ is the radial distance, and
$\vartheta, \varphi$, as usual, are the polar angles.\\
2. The scale factor, the pressure and the energy density are denoted in FRW and in EFRW by $R, p, e$.\\
3. In EFRW let $t^*: = inf I = - \infty$ and $t^\sharp: = sup\; I = \infty$ and in FRW let $t^*: = inf I^\star = 0$ whereas $t^\sharp: = sup\; I^\star$
is finite or infinite.\\
4. The function $J$ is defined by

\begin{equation} \label{40}
{J} (t, t') = \int\limits^{t}_{t'} |R (s)|^{-1} ds      
\end{equation}

for each pair $t, t'$ for which the integral exists.

Then we obtain the following

{\bf Consequences:} 1. If $t, t'> 0$ or if $t, t' < 0$ the integrand in (\ref{40}) is bounded. Hence ${J} (t, t')$ exists and
${J} (t, t') = {J} ( -t', -t)$.\\
2. In order to evaluate ${J} (t, t')$ for $t, t'>0$ or for $t, t' < 0$ one need not solve the field equations. It suffices to know the
constitutive equation $e = {\hat e} (R)$ and the two values $R(t)$ and $R(t')$.

The proof runs as follows. First, let $\dot{R} (s) >0$ for $t' \leq s \leq t$. Then the equation, $\sigma = R(s)$ is solvable for $s$ so that $\sigma$
can be used as intergration variable. Therefore

\begin{equation} \label{41}
J (t, t') = \int\limits^{R(t)}_{R(t')} | \sigma|^{-1} \dot{R}^{-1} d \sigma.
\end{equation}

Inserting (\ref{10}) into (\ref{41}) one obtains 

\begin{equation} \label{42}
J (t, t') = 3^{\frac{1}{2}} \int\limits^{R(t)}_{R(t')} | \sigma|^{-1}
(\kappa_0 \hat{e}(\sigma) \sigma^2 + \Lambda_0 \sigma^2 - 3 k)^{-\frac{1}{2}} d \sigma.
\end{equation}

If $\dot{R}$ changes sign in $[t', t]$ then this interval is the union of subintervals for which (\ref{42}) holds,
and $J (t, t')$ is a sum which adds up to (\ref{42}).

With the help of (\ref{42}) now it is easy to check if $J (t, t')$ exists for $t'<0<t$.

{\bf Proposition:} If matter is asymptotically radiation dominated for $t, t' \rightarrow 0$, then $J (t, t')$ 
exists for $t' < 0 < t$.

{\bf Proof:} It suffices to prove $J (t, 0) < \infty$ for $t >0$. By supposition $\hat{e} (\sigma) = B \sigma^{-4}$ with $B>0$, and
$0 \leq \sigma \leq R(\bar{t})$ for some $\bar{t} >0$. Now let

\begin{equation} \nonumber
F(\sigma): = \kappa_0 B + \Lambda_0 \sigma^4 - k \sigma^2.
\end{equation}

Then there is $m>0$ such that $F(\sigma) > m$ for sufficiently small $\sigma$, i.e. for sufficiently 
small $\bar{t}$. The integrand in
$J (\bar{t}, 0)$ according to (\ref{42}) is $F(\sigma)^{-\frac{1}{2}}$. Hence

\begin{equation} \label{43}
J (\bar t, 0) \leq 3^{\frac{1}{2}} m^{- \frac{1}{2}}  R(\bar{t}) < \infty.
\end{equation}

Because of $J (t, 0) = J (t, \bar{t}) + J (\bar{t}, 0)$ the propositions holds.

{\bf 3.1.2 Particle Horizons}\\ 
{\bf 3.1.2.1:} Since ponderable matter is assumed to be at rest in $N_k$ a particle can be identified
with its position in $N_k$. Consequently, the trajectory $\gamma_a$ of a particle $a \in N_k$ is given by
 $\gamma_a (t) = (a,t)$ with $t \in I$ or $I^\star$ respectively, and the worldline of $a$ is $W_a: = ran \; \gamma_a$.
Then the concept of a particle horizon is defined as follows:

{\bf Definition:} The particle horizon ${\mathcal H} (b,t)$ of an observer $b$ at time $t$ is the boundary between the
set of particles ${\mathcal K} (b,t)$ which can send signals arriving at $b$ up to time $t$, and the set of particles which can not.

If ${\mathcal H} (b,t) = \emptyset$, it is usual to say that $b$ does not have a particle horizon at time $t$.

{\bf 3.1.2.2:} The homogeneity and isotropy of the manifolds considered in EFRW and FRW allow to formulate a simple
criterion for the existence of a particle horizon as follows:

{\bf Proposition: }1. Case $k = 0, -1$. The particle horizon ${\mathcal H} (b,t)$ of an observer $b$ at time $t > t^\star$ exists
exactly if ${J} (t, t^*) < \infty$.\\
2. Case $k = 1$. The particle horizon ${\mathcal H} (b,t)$  of an observer $b$ at time $t > t^\ast$ exists exactly if
${J} (t, t^*) < \pi$.\\
3. In both theses cases the particle horizon is given by 


\begin{equation} \nonumber
{\mathcal {H}} (b,t)  = \partial {\mathcal {K}} (b,t) 
\end{equation}

and
\begin{equation}\label{44}
{\mathcal {K}} (b,t) = \{ (\chi, \vartheta, \varphi): 0 \leq \vartheta \leq \pi, 0 \leq 2 \pi, \chi \leq {J} (t, t^*) \}.
\end{equation}
The proof can be extracted from the literature, especially from (\cite{Rind1}, \cite{Rind2}, \cite{Goe1}).

Immediately from the proposition one obtains the following

{\bf Consequences: }1. If ${\mathcal H} (b,t) = \emptyset$ and if $t' > t$ then also ${\mathcal H} (b,t') = \emptyset$, because
${J} (t',t^\ast) \geq {J} (t, t^*)$.\\
2. If ${\mathcal H} (b,t) \not= \emptyset$ and if $t'' < t$ then also ${\mathcal H} (b,t'') \not= \emptyset$, because ${J} (t'', t^\ast) < {J} (t, t^*)$.

The physical significance of the concept of a horizon is underlined by the following

{\bf Proposition:} $b_2 \in {\mathcal K} (b_1, t)$ if and only if $b_1 \in {\mathcal K} (b_2, t)$.

The proof can be read off from (\ref{44}) if one takes into account that if $b_1 = (\chi_1, \vartheta_1, \varphi_1)$
and $b_2 = (\chi_2, \vartheta_2, \varphi_2)$ in their respective coordinate systems, we have $\chi_1 = \chi_2$, because
$\chi_1$ is the distance between $b_2$ and $b_1$ in the Riemannian space $(N_k, h_k), k = 0, \pm 1$, and $\chi_2$
is the distance between $b_1$ and $b_2$.

The proposition can be physically interpreted: Let $b$ be a particle. Then $b$ is causally affected by all particles $a$
out of the horizon of $b$ at time $t$ and its interior, and $b$ affects causally up to time $t$ all particles out of its horizon 
and its interior (at $t$). Thus, if there are no horizons all particles are in causal contact with each other.

{\bf 3.1.3 Event Horizons}\\
{\bf 3.1.3.1:} In this Section again the notation of Section 3.1.1 is used especially 
$t^\sharp: = \sup \;I$ or $t^\sharp: = \sup I^\star $.
Then the concept of an event horizon is given by the following

{\bf Definition:} The event horizon ${\mathcal E} (b,t)$ of an observer $b$ at time $t$ is the boundary of the set ${\mathcal L} (b,t)$ of events
$(a, t) \in N_k \times \{t\}$ from which signals reach $b$ up to time $t^\sharp$ and of the set of events $(a', t) \in N_k \times \{t\}$ from which
signals cannot reach $b$.

{\bf 3.1.3.2.:} For the existence of event horizons also a simple criterion exists.

{\bf Proposition: }1. Case $ k=0, -1$. The event horizon  ${\mathcal E} (b,t)$ of an observer $b$ at time $t < t^\sharp$ exists exactly if
${J} (t^\sharp, t) < \infty$.\\
2. Case $k=1$. The event horizon ${\mathcal {E}} (b,t)$ of an observer $b$ at time $t < t^\sharp$ exists exactly, if
$J (t^\sharp, t) < \pi$.\\
3. In both these cases the event horizon is given by 

\begin{equation} \nonumber
{\mathcal {E}} (b,t)  = \partial {\mathcal {L}} (b,t) 
\end{equation}

and
\begin{equation}\label{45}
{\mathcal {L}} (b,t) = \{ (\chi, \vartheta, \varphi, t): 0 \leq \vartheta \leq \pi, 0 \leq \varphi \leq 2 \pi, \chi \leq {J} (t^\sharp, t) \}.
\end{equation}

Again there are some immediate

{\bf Consequences: }1. If ${\mathcal E} (b,t)= \emptyset$  and if $t' < t$ then also  ${\mathcal E} (b,t')= \emptyset$ because 
${J} (t^\sharp, t') \geq {J} (t^\sharp, t)$.\\
2. If  ${\mathcal E} (b,t) \not= \emptyset $ and if $t'' > t$ then also  ${\mathcal E} (b,t'') \not= \emptyset$ because   
${J} (t^\sharp, t'') < {J} (t^\sharp, t)$.

{\bf 3.1.4:} For a discussion of the horizon structure of FRW I refer to the cited literature (c.f. e.g. \cite{Kolb1}, \cite{Rind1}, \cite{Haw1},
\cite{Goe1}). For $k = 0, -1$, the existence of particle horizons follows directly from the proposition of Section 3.1.1.

\subsection{Horizons in EFRW}

{\bf 3.2.1:} One of the characteristics of EFRW is the symmetry with respect to time. Again the notation
of Section 3.1.1 is used. Then we obtain the following

{\bf Proposition:}
A particle horizon for an observer $b$ at time $t$ exists exactly if an event horizon for $b$ exists at time $-t$.

{\bf Proof:} By definition of ${J}$ the equation ${J} (t,t') = {J} (-t', -t)$ holds, thus also if
$t' = - \infty$. Using the criteria of 3.1.2 and 3.1.3 the proposition is seen to hold.

Hence for a discussion of horizons in EFRW we need to calculate only one of the integrals ${J} (-t,- \infty)$
or ${J} (\infty, t)$. In Subsection 2.5.3.3 it was shown that the models of EFRW with changing signs of 
$\dot R (t)$ can be periodically extended to the interval $I = \mathbb R$ in a natural way.
Hence all models of EFRW have the same time axis $I = \mathbb R$.

{\bf 3.2.2 Periodic Models}\\
This case is the simplest one in EFRW. For each periodic model we obtain the following

{\bf Proposition:} There is neither a particle horizon nor an event horizon for any particle $b \in N_k, k = 0, \pm 1$ at any
$t \in I = \mathbb R$.

{\bf Proof:} Let $T$ be the length of a period, and let $t$ be any time. Then there is a time $t'$ such that $0 <t' - t < T$ and
$J (t',t) >0$. Hence $J (t',t) = J (t' + n T, t + n T) $ and 

 \begin{equation} \label{46}
J (\infty,t) \geq \sum\limits^\infty_{n = 0}
J (t' + n T, t + n T) = \infty. 
\end{equation}

{\bf 3.2.3 Nonperiodic Models}\\
{\bf 3.2.3.1:} In this section models are considered for which $\dot R (t) > 0$ for all $t \in I = \mathbb R$.

Moreover, it is assumed that there is a time $t_1 >0$ such that for $t >t_1$ space--time is asymptotically matter dominated and matter is dust.
Therefore, one obtains from (\ref{12}) the constitutive equations:

\begin{equation} \label{47}
p = 0, \quad  e = A |R|^{-3}
\end{equation}

with $A = e(t_0) |R (t_0)|^3$. Then $e (t) = e(-t)$.

As usual, it is supposed that the universe near big bang is radiation dominated.

Finally, in what follows we will study only such models for which $R(t) \rightarrow \infty$ holds
 if $t \rightarrow \infty$. (For the other cases
cf. e.g. \cite{Rind1}).

These suppositions have two

{\bf Consequences: }1. For $t > t_1$ equation (\ref{42}) with (\ref{47}) reads:

\begin{equation} \label{48}
J (\infty, t) = 3^{\frac{1}{2}} \int\limits^\infty_{R(t)} \sigma^{-1} (\kappa_0 A \sigma^{-1} + \Lambda_0 \sigma^2 - 3k)
^{-\frac{1}{2}} d \sigma.
\end{equation}

2. If $\Lambda_0 < 0$, the bracket in (\ref{48}) becomes negative for large $\sigma$. Therefore, $\Lambda_0 < 0$ is not compatible
with the above assumptions so that we have to study only the cases $\Lambda_0 = 0$ and $\Lambda_0 >0$.

{\bf 3.2.3.2:} Let $\Lambda_0 = 0$ and let us assume that the above suppositions are valid. Then necessarily $k = 0, -1$, and one obtains
the following result.

{\bf Proposition:} There is neither a particle horizon nor an event horizon for any particle $b \in N_k, k = 0, -1$ at any time 
$t \in I = \mathbb R$.

{\bf Proof:} Let us consider $J (\infty, t')$ for a time $t' > t_1 >0$ out of the dust era of matter. Then (\ref{48}) can be applied
and the variable $\sigma$ in (\ref{48}) obeys the inequality $\sigma \geq R(t') > 0$. Therefore
$\kappa_0 A \sigma^{-1} + 3 |k| \leq \kappa_0 A R (t')^{-1} + 3 |k| = : N (t')^{-2}$ where $N(t') >0$. Hence for $t' >t_1$  we have

\begin{equation} \label{49}
J (\infty, t') \geq 3^{\frac{1}{2}} N(t') \int\limits^\infty_{R(t') }\sigma^{-1} d \sigma = \infty.
\end{equation}

For each other time $t$ the equation

\begin{equation} \label{50}
J (\infty, t) = J (\infty, t') + J (t', t) = J (-t, -\infty)
\end{equation}

holds. Then by the critera of the Sections 3.1.2.2 and 3.1.3.2 the proposition follows at once.

{\bf Remark:} This result has consequences for what is called the horizon problem. Strictly speaking, where there are no horizons
there cannot be a {\sl {horizon}} problem. At least, if there is a problem it does not make sense to call it ''horizon'' problem in this case.

{\bf 3.2.3.3.:} Now let $\Lambda_0 > 0$ and $k = 0, -1$. Then with the above suppositions we obtain the following

{\bf Proposition:} For each particle $b \in N_k, k= 0, -1$, and each time $t \in I = \mathbb R$ there is both a particle horizon and an
event horizon.

{\bf Proof:} It follows from (\ref{48}) that for $t'$ in the dust era

\begin{equation} \label{51}
\begin{array} {cc}
J (\infty, t') = 3^{\frac{1}{2}}\int\limits^\infty_{R(t') }\sigma^{-2} 
(\Lambda_0 + \kappa_0 A \sigma^{-3} + 3|k| \sigma^{-2} )^{-\frac{1}{2}} d \sigma \\
\leq 3^{\frac{1}{2}} \Lambda_0^{-\frac{1}{2}} \int\limits^\infty_{R(t')} \sigma^{-2} d \sigma  < \infty.
\end{array}
\end{equation}

Hence with (\ref{50}) the proposition is seen to hold.

{\bf 3.2.3.4.:} Let us finally consider the case $\Lambda_0 > 0$ and $k = 1$. Then one can show that the above
conditions are only satisfiable if $\Lambda_0$ is restricted by the stronger condition $\Lambda_0 > \beta^{-2}$ where
$\beta = \frac{1}{2} \kappa_0 A > 0$ (cf. e.g. (\cite{Rind1}). But, in order to discuss the present case completely we had to specify
the constitutive equations between matter domination and radiation domination, which is outside the scope of this paper. Thus,
we confine ourselves to the simplest achievable result.

{\bf Proposition:} For each $\Lambda_0 > \beta^{-2}$ there is a time $t_2 \geq t_1 > 0$ such that for each particle
$b \in N_1$ a particle horizon exists for $t' < -t_2$ and an event horizon for $t > t_2$.

{\bf Proof:} It follows from (\ref{48}) that for each $\Lambda_0 > \beta^{-2}$ the inequality

\begin{equation} \label{52}
J (\infty, t) < 3^{\frac{1}{2}}\int\limits^\infty_{R(t) } h (\sigma)^{-\frac{1}{2}} d \sigma = : X (t)
\end{equation}

holds with

\begin{equation} \label{53}
\begin{array} {ll}
h (\sigma) &= 2 \beta \sigma + \beta^{-2} \sigma^{4} - 3 \sigma^{2}\\
&= \beta^{-2} \sigma (\sigma + 2 \beta) (\sigma - \beta)^2.
\end{array}
\end{equation}

Therefore

\begin{equation} \label{54}
X (t) = 3^{\frac{1}{2}} \beta \int\limits^\infty_{R(t) } (\sigma^{2} + 2 \beta \sigma)^{-\frac{1}{2}} |\sigma - \beta|^{-1} d \sigma.
\end{equation}

The last integral is elementarily solvable. It turns out that it is finite, hence 

\begin{equation} \nonumber
\lim_{t \rightarrow \infty} X (t) = 0.
\end{equation}

Thus, there is a time $t_2$ such that $X (t_2) \leq \pi$.

{\bf 3.2.4}:
The results of the Sections 3.2.2 and 3.2.3 can be summarized as follows:

Each particle of a universe in EFRW is at each time in causal contact with all other particles if the universe is either periodic or aperiodic and
$\Lambda_0 = 0$. This is an immediate consequence of the fact that there are no horizons in these universes.
Some remarks on the so-called horizon problem can be found in Section 4.4.\\

\section{Discussion of the Results}

\subsection{How singular is big crunch bang in EFRW?}
{\bf 4.1.1:}
In order to get a sound answer to this question one has to say some words about the notion of a singularity as far as necessary for the 
purposes of this paper. The exact definition of a singularity in a space-time manifold can be found e.g. in the classic monograph of Hawking and Ellis \cite {Haw1} 
and in some other texts (cf. \cite {pen1, ger1,tip}). Besides this exactly defined notion other, more intuitiv concepts of a singularity are used. 
They are connected with terms like undefiniteness, non-differentiability, divergence etc. In what follows I call these intuitive concepts 
weak singularities.

{\bf 4.1.2:}
Let us first consider the manifold $M$ as it is introduced in Section 2.5.1.

By definition, $M = N_k \times I, k = 0, \pm 1$ where $N_k$ does not depend on time. Hence, 
there is nothing special with $M$. (The same is the case for the manifold $M^\star$ of FRW.)

The kinematics of matter in $M$ is given by the trajectories $\gamma_a$ of particles $a \in N_k$ which are defined by 
$\gamma_a (\tau) = (a, \tau), \tau \in I$. For each $a \in N_k$ the curve $\gamma_a$ is a timelike affine geodesic, i.e. the particles
$a \in N_k$ are freely falling. The set of all trajectories is a smooth congruence on $ M$ which is inextendible with respect to the non-affine 
parameter $\tau$. There is no singularity in the weak sense.

{\bf 4.1.3:}
Let us now consider a lightlike signal between two points $ p_1 = (a_1, \tau_1)$ and $ p_2 = (a_2, \tau_2)$ of $M$, and let $(\chi, \vartheta, \phi)$ be 
defined as in Notation 1. of  Section 3.1.1 such that $ a_1 $ is the origin of the coordinate system. Moreover let be $a_2 = (\chi_2, \vartheta, \phi)$ and 
let $\sigma $ be defined by 

\begin{equation} \label{55}
\sigma(\tau) = (\psi(\tau), \vartheta, \varphi, \tau) 
\end{equation}

where
\begin{equation} \label{56}
\psi(\tau)  =  \int\limits^{\tau}_{\tau_1}\frac {\dot f( \lambda)}{\left|S(\lambda)\right|} d\lambda  =  J (f(\tau),f(\tau_1))
\end{equation}

with $S$ and $f$ as in Section 2.5.1. Then the following result holds.\\
{\bf{Proposition:}} 1. $\sigma$ is a lightlike curve between $p_1$ and $p_2$ if $\chi_2 = \psi(\tau_2)$\\
2. $\sigma $ is defined for all $\tau_1, \tau_2 \in \mathbb{R}$ and is of class $C^2$ if matter is radiation dominated 
for $\tau\rightarrow 0$ and if the relations (\ref{30}) hold.\\
3. $ \sigma$ is an affine geodesic which is defined for all $\tau \in \mathbb{R}$. Hence it is inextendible. \\
The proof is simple but lengthy, only the case $k = 1$ needs some care.\\
Since each lightlike signal can be brought into the special form (\ref{55}) we have obtained the complete null-geodesic structure. 
It is of class $C^2$ 
with respect to the non-affine parameter $\tau$. Hence, there is also no singularity in the weak sense.

{\bf 4.1.4:}
Things change if one takes the metric $g$ into account. As already remarked in Subsection 2.5.3.1 the metric becomes singular in the
sense that $g = 0$ so that the invers of $g$, i.e. the contravariant tensor $g^\sharp$, does not exist at $\tau = 0$. However, this property
does not affect $M$ as a manifold. Rather the measurements of proper length and proper times degenerate at $\tau = 0$, and this
singularity is not removable within EFRW. But, in all models and applications of the theory the metric does not show up as a tensor.
It suffices to know the functions $S$ and $\dot f$ in (\ref{36}). Therefore the singularity of $g$ boils down to the two relations
$S (0) = 0$ and $\dot f (0) = 0$. The theoretical fact that $g=0$ for $\tau = 0$, recedes to the background because it does not affect
the practical work.

{\bf 4.1.5:}
It is often believed that the most characteristic feature of big bang or big crunch bang is the divergence of the constitutive quantities 
$\bar p$ and $\bar e$ which is
caused by $S (\tau) \rightarrow 0$ for $\tau \rightarrow 0$. But this is not the case, this kind of singularity can be removed totally!

As we have seen in the Sections 2.3 and 2.4.3 the quantities $\bar p$ and $\bar e$ are not suitable for a theoretical description of big crunch
bang. Following the result of Section 2.4.2 the quantities $\psi = \bar{p}^{-1}$ and $\omega = \bar{e}^{-1}$ are suitable for this purpose. Introducing
such ''suitable quantities'' instead of $\bar p$ and $\bar e$ looks like a dirty trick. But this is not the case as I will demonstrate by three arguments.

1. Let us assume that big crunch bang exists as a natural phenomenon. Then its existence is not touched by its theoretical description with the 
help of suitable quantities. The contrary is the case. An extraordinary phenomenon needs special treatment. The only criterion is that
the latter is unobjectionable with respect to physics and mathematics.

2. If a singularity of a certain kind is thought to be essentially determined by the divergence of some physical quantities then in each
field theory an arbitrary number of singularities can be constructed as follows. Let $x$ be a point of a space-time, and let $A_1, \ldots, A_n$
be fields of the theory. Now introduce new fields by $B_j = (A_j -A_j (x))^{-1}$, $j = 1, \ldots, n$, then $B_j$ diverges at $x$. Hence $B_1, \ldots, B_n$ are not suitable
to describe the physical situation at $x$. 

3. The ''dirty'' trick, i.e. introducing suitable quantities, was already used when FRW was derived from General Relativity specialised to 
homogeneous and isotropic spaces. There the primary quantity is scalar curvative $\hat C$. But it is more adequate to introduce the scale factor
$K$ by $\hat C = k K^{-2}$.

These arguments show that introducing suitable quantities to remove divergent terms is far from being a dirty trick.

{\bf 4.1.6:}
Summing up the results of 4.1.2 to 5, we see that in EFRW the kinematic and thermodynamic stuctures of matter as well as the structure of lightlike geodesics are smooth and inextendible if one uses the new cosmic time $\tau$. These properties of EFRW are very satisfactory. But, the question remains if the Lorentz manifold $(M,g)$ is singularity-free in the sense of Hawking, Ellis 
et al \cite{Haw1, pen1, ger1, tip} or not. I think it is not! The reason is that the concept of a singularity (in this strong sense) depends on the use of affine (or generalized affine) parameters, and this has the consequence that proper time comes into play. Thus we are confronted with the same situation as in FRW. But, as we will see in Section 4.2, proper time loses its physical meaning in the vicinity of big crunch bang, because standard clocks indicating proper time are not constructible (cf.\cite{Per1}). Hence the analysis of a singularity of $(M,g)$ in the strong sense reveals only a mathematical property of EFRW which is not physically relevant because it does not affect the physical properties of matter and lightlike signals. It is nothing but attributing the term singular to a certain hyperplane. This guarantees the physical applicability of EFRW. The situation is similar to that described in Section 4.1.5. The choice of unsuitable quantities can produce singularities in a wide sense. In the present case proper time is not suitable.

\subsection{The clock problem in EFRW}

In Section 2.4 we have seen that there are mathematical reasons to work with a new cosmic time scale $\tau$ instead of the time scale $t$ which is
the proper time of the particles of ponderable matter. In this section I will complement the considerations of 2.4 by a physical argument in
favour of $\tau$.

Proper time was first introduced as a mathematical term. Later on the problem was studied whether it is possible to define proper time as that time
which is indicated by a special class of clocks, the so-called standard clocks, which are constructible (in principle!) solely by geometrical means.
Such clocks exist (cf. \cite{Per1}, \cite{Kun1}, \cite{Mar1}). They indicate the proper time of a particle which shows up in the
construction of a standard clock. Then by experience we know that the time scales of standard clocks and of atomic clocks coincide at the present
era of earth. The latter clocks are only constructible with the help of quantum mechanical means.

Within dense matter as in the universe near big crunch bang the geometric construction principles of standard clocks lose their reference to 
reality completely, because the freely falling test particles needed for their construction do not exist (cf. \cite {Per1}). Proper, i.e. standard time then is nothing but a theoretical concept. This is a strong hint that it should be replaced by
another time scale. This was done in Section 2.4 by introducing the new cosmic time scale $\tau$. But, up to now it is an open problem whether
there are constructive principles based on quantum theory for the clocks indicating time $\tau$. Nevertheless, it should be mentioned again that
outside a neighbourhood of $\tau = 0$ the proper time scale $t = f (\tau)$ is appropriate, at least for practical purposes.

\subsection{The scale factor}

At a first glance, the property $S (\tau) < 0$ for $\tau <0$ could look strange. But one should be aware that even in FRW the supposition
$K (t) >0$ for $t >0$ is nothing but a comfortable convention. One could also work with the convention $K (t) < 0$ for $t >0$. Also in EFRW
there does not occur anything special. The reason is that many physical quantities, e.g. curvature and metric, depend only on $S^2$ or $|S|$.
Other quantities, e.g. the Hubble parameter, have to be redefined properly.

\subsection{The horizon problem}

The results of Chapter 3 formulated in Section 3.2.4 can also be expressed in the following form.

A model of EFRW does not have horizons exactly if it is either periodic or aperiodic with $\Lambda_0 = 0$. Then, in the latter case necessarily
$k = 0, -1$. Moreover, if $k = 0, -1$ periodicity occurs if and only if $\Lambda_0 <0$.

This result suggests complementing EFRW with the additional condition

{\bf AC:} The solutions $S, \psi, \omega$ are either periodic or aperiodic with $\Lambda_0 = 0$.

In order to discuss the consequences of the above result with respect to the horizon problem let us first give a short characterization of it. \\
If a relativistic theory, denoted $ \Psi$, has symmetries, e.g. homogeneity and isotropy, two questions arise.\\
1. Are there horizons in $\Psi$?\\
2. If there are horizons in $\Psi$, why is it that the manifold $M$ of $\Psi$ exhibits symmetries, i.e. why have causally separeted regions of $M$ 
the same or very similar properties?\\
These two questions constitute what is called horizon problem in the present context. Since in EFRW plus AC the answer to the first question is "No", so 
that the second question is empty, the horizon problem has been solved.

But one should be aware that there is a problem behind the horizon problem which has not vanished with vanishing horizons. For, the nonexistence
of horizons only guarantees causal contact for each particle with all other particles in the world.
Intuitively, this is a necessary condition for the large scale homogeneity and isotropy. But is it also sufficient? Hence, the horizon problem has a successor, the problem of large scale stability of universes with full causal contact of all particles. As far as I can see this
problem is unsolved. However, calling it also "horizon problem" is abuse of language. This is because the term "horizon problem"
has become an attribute of elements of the null set.

\subsection{Outlook}

Looking back at EFRW as it is presented in Section 2.5 one may notice that the theory contains unspecified elements, the constitutive quantities
$p$ and $e$, and the cosmological constant $\Lambda_0$. Only occasionally some specifications were introduced. It was assumed in Chapter 3
that the universe near big crunch bang is asymptotically radiation dominated and that it is asymptotically matter dominated and
$p = 0$ (i.e. matter is dust) for sufficiently large $|\tau|$. 
Moreover, in Section 4.4 the condition AC was introduced which amounts roughly to $\Lambda_0 \leq 0$. This was done on purpose! It was not
intended to study specified cosmological scenarios, rather to formulate a frame theory for possible scenarios.

In this way EFRW is open to various combinations, especially all such scenarios which are used in FRW, like inflation, variable speed of light 
etc. (cf. e.g. \cite{Kolb1}, \cite{Alb1}, \cite{Gas1}) and the literature quoted there.)

But EFRW plus AC has the advantage that it avoids some severe shortcomings of FRW as it was demonstrated in this paper. This is a strong hint to cast out 
all models with open space-times and positve cosmological constants.\\

{\bf Acknowledgement}

I want to thank my colleagues Mr. Gerhard Lessner for valuable discussions and critical remarks and Mr. Rolf Breuer for correcting my Englisch.

\bibliographystyle{my-h-elsevier}

\end{document}